\documentclass[12pt,epsf,amssymb]{article}

\usepackage{graphicx}
\usepackage{amsfonts}
\usepackage{amsmath}
\usepackage[usenames]{color}
\usepackage{amssymb}
\usepackage[mathscr]{eucal}
\usepackage{url}

\usepackage[all]{xy}

\def\N{\mathbb{N}}
\def\Z{\mathbb{Z}}

\def\R{\mathbb{R}}
\def\C{\mathbb{C}}
\def\P{\mathbb{P}}

\usepackage{here}

\newcommand{\parren}[1]{\left( #1 \right)}

\newcommand{\sqbra}[1]{\left[ #1 \right]}

\newcommand{\pdiff}[3][]{\dfrac{\partial^{#1}{#2}}{\partial{#3}^{#1}}}

\newcommand{\tensor}{\otimes}

\newcommand{\Tr}{\operatorname{Tr}}

\usepackage{amsthm}
\theoremstyle{definition}

\begin{document}

\baselineskip 0.6cm
\newcommand{\vev}[1]{ \left\langle {#1} \right\rangle }
\newcommand{\bra}[1]{ \langle {#1} | }
\newcommand{\ket}[1]{ | {#1} \rangle }
\newcommand{\Dsl}{\mbox{\ooalign{\hfil/\hfil\crcr$D$}}}
\newcommand{\nequiv}{\mbox{\ooalign{\hfil/\hfil\crcr$\equiv$}}}
\newcommand{\nsupset}{\mbox{\ooalign{\hfil/\hfil\crcr$\supset$}}}
\newcommand{\nni}{\mbox{\ooalign{\hfil/\hfil\crcr$\ni$}}}
\newcommand{\nin}{\mbox{\ooalign{\hfil/\hfil\crcr$\in$}}}
\newcommand{\Slash}[1]{{\ooalign{\hfil/\hfil\crcr$#1$}}}
\newcommand{\EV}{ {\rm eV} }
\newcommand{\KEV}{ {\rm keV} }
\newcommand{\MEV}{ {\rm MeV} }
\newcommand{\GEV}{ {\rm GeV} }
\newcommand{\TEV}{ {\rm TeV} }

\def\diag{\mathop{\rm diag}\nolimits}
\def\tr{\mathop{\rm tr}}

\def\Spin{\mathop{\rm Spin}}
\def\SO{\mathop{\rm SO}}
\def\SU{\mathop{\rm SU}}
\def\U{\mathop{\rm U}}
\def\Sp{\mathop{\rm Sp}}
\def\SL{\mathop{\rm SL}}

\def\change#1#2{{\color{blue}#1}{\color{red} [#2]}\color{black}\hbox{}}

\begin{titlepage}

\begin{flushright}
IPMU20-0049
\end{flushright}

\vskip 1cm
\begin{center}

{\large \bf Witten Anomaly in 4d Heterotic Compactificaitons with ${\cal N}=2$ Supersymmetry}

 \vskip 1.2cm

Yuichi Enoki, Yotaro Sato and Taizan Watari,

 \vskip 0.4cm

  {\it Kavli Institute for the Physics and Mathematics of the Universe (WPI),
     the University of Tokyo, Kashiwa-no-ha 5-1-5, 277-8583, Japan}

 \vskip 1.5cm

 \abstract{
    We showed that there is no $SU(2)$ Witten anomaly in a large class of 4d ${\cal N}=2$ supersymmetric Heterotic string compactifications. The consistency conditions 
    we consider are the modularity of the new supersymmetric index, the integrality of BPS indices, and the discrete Peccei--Quinn shift symmetries. We also found an example where these conditions are not sufficient to show that the theory is anomaly free. This suggests that there are more conditions in the worldsheet SCFT essential for consistent string compactifications.
 }

\end{center}
\end{titlepage}


\newpage
\section{Introduction}

Theoretically consistent string vacua are believed to yield 
low-energy effective field theories without a theoretical inconsistency. 
If we find a low-energy inconsistency in a string vacuum, that is taken 
as a hint of some more theoretical consistency conditions of string theory.
Gauge and gravitational anomalies in the effective theories in closed string 
vacua are known to cancel due to the modular invariance of the CFT \cite{SW}.
The same class of anomalies in the effective theories in open string vacua 
cancel locally because of the anomaly inflow mechanism \cite{inflow,brane-CS},
and they cancel globally because of the Bianchi identity of Ramond--Ramond 
fields.  

In this article, we address the $SU(2)$ Witten anomalies \cite{W} of the 4d $\mathcal{N}=2$ supersymmetric effective theories given by Heterotic compactifications.
The organization of this article is as follows.

Section 2 is a short review of basic facts about the $SU(2)$ Witten anomaly.
Section 3 explains our setup (including the choice of $SU(2)$ gauge symmetry) and some assumptions for the Heterotic compactifications.

In section 4, we show that the Witten anomaly indeed vanishes for the lattice\footnote{
In our setup, $\widetilde{\Lambda}_S = U[-1] \oplus \Lambda_S$ is the charge lattice of chiral free bosons in the internal worldsheet CFT.
The charge vector $v_0 \in \widetilde{\Lambda}_S^\vee$ specifies  the choice of the $SU(2)$ gauge symmetry.
See section 3 for more details.
} $\Lambda_S$ and the charge vector $v_0$,
if $\Lambda_S$ has a decomposition $U \oplus W$ for some even negative definite lattice $W$ and the charge vector $v_0$ stays within $W^\vee$.
The derivation described in section 4.1 exploits 
\begin{itemize}
    \item the modularity of the generating functions\footnote{
    See section 3 for the definitions for the modular form $\Phi$. See also 5.1 for the modular form $\Psi$.
    } $\Phi$ and $\Psi$ for BPS indices, defined by the internal worldsheet CFT language, 
    \item the integrality of BPS indices, and
    \item the integrality of some coefficients of the leading terms in the prepotential for the effective theory, which follows from the discrete Peccei--Quinn shift symmetries.
\end{itemize}
Section 4.2 shows that for some choices of the lattice $\Lambda_S$ and the charge vector $v_0$, the cancellation of the Witten anomaly follows only by the modularity (the first condition).

Section 5 treats an example of $\Lambda_S$ and $v_0$ for which one cannot show that the Witten anomaly always vanishes only from these conditions listed above.
This means that we are missing some other important consistency conditions for the string compactifications.

In the appendix, we review what the discrete Peccei--Quinn shift symmetries imply for the leading coefficients of the prepotential.

\section{$SU(2)$ Witten Anomaly}

Since 4d $\mathcal{N}=2$ supersymmetric theories are non-chiral, they are free from perturbative triangle anomalies. However there can be nonperturbative anomalies, called Witten anomalies \cite{W}.
In $\mathcal{N}=2$ $Sp(k)$ gauge theories, a half-hypermultiplet massless matter in the representation $R$ with half-integral Dynkin index $T(R)$ produces Witten anomaly.
In the case of $SU(2)$ gauge theory, the possibly anomalous representations are those with the spin
\begin{equation}
    \label{anomalous-j}
    j = \frac{1}{2} + 2l, \qquad l=0,1,2,\dots.
\end{equation}
The theory is free from Witten anomaly if and only if the total number of massless $SU(2)$ half-hypermultiplets in such representations is even, i.e. 
\begin{align}
    \mathrm{WA} := \sum_{j \in \frac{1}{2} + 2 \Z_{\geq 0}} N(R_j) \equiv 0 \bmod 2,
\end{align}
where $N(R_j)$ is the number of massless half-hypermultiplets in the spin-$j$ representation $R_j$.
This quantity can also be written by $N_j$, the number of massless half-hypermultiplets with the $2j$ units of the fundamental weight 
(= $SU(2)$ Cartan charge $j$):
\begin{align}
    \label{eq:WA-vanish}
    \mathrm{WA}
    = \sum_{j \in \frac{1}{2} + 2 \Z_{\geq 0}} (N_j - N_{j+1})
    \equiv \sum_{j \in \frac{1}{2} + \Z_{\geq 0}} N_j    
    \bmod 2.
\end{align}
Note that this summation may run over those states in $SU(2)$ full-hypermultiplets because they appear in pairs and have no contributions modulo two. Moreover, if the theory has other $U(1)$ gauge symmetries, only neutral states under those $U(1)$'s contribute to (\ref{eq:WA-vanish}) because $U(1)$-charged matter has its conjugate and this pair contributes 2.

\section{Heterotic compactification and $SU(2)$ enhancement}
\label{sec:set-up}

Let us consider a Heterotic compactification with 4d $\mathcal{N}=2$ 
supersymmetry; we assume that it is without an $\R^{3,1}$-filling NS5-brane 
and other solitons of similar kinds. The internal $(c,\tilde{c}) = (22,9)$ 
CFT in the NSR formalism with $(0,2)$ worldsheet supersymmetry for such 
a compactification is characterized by the conditions in \cite{BDetal}.
Let $\widetilde{\Lambda}_S$ denote the charge lattice of chiral free bosons 
that appear in the CFT \cite{AFGNT-pert, HM}, 
$G_S := \widetilde{\Lambda}_S^\vee/\widetilde{\Lambda}_S$ its discriminant 
group, and $\rho := {\rm rank}(\Lambda_S)$.
We assume
that the lattice $\widetilde{\Lambda}_S$ has a primitive embedding into 
$U^{\oplus 4}\oplus E_8[-1]^{\oplus 2}$, where $U$ is the even unimodular lattice 
of signature $(1,1)$, and a structure\footnote{
In the language of Type IIA dual, this structure corresponds to the 0-form and 4-form cohomology classes of the K3-fiber of the internal Calabi--Yau 
threefold staying distinct from the 2-form cohomology classes of the fiber.
} %
$\widetilde{\Lambda}_S = U[-1] \oplus \Lambda_S$.

When a charge vector $v_0 \in \widetilde{\Lambda}_S^\vee$ satisfies the 
following conditions, 
\begin{align}
    \frac{(v_0,v_0)}{2} = -\frac{1}{k}, \qquad
    kv_0 = 0 \in G_S, \quad {}^\exists k \in \Z_{>0}, 
    \label{eq:enhancement-condition}
\end{align}
and\footnote{
$(v_0,v_0)$ or $v_0^2$ denotes the square in the intersection form of 
$\widetilde{\Lambda}_S$.
} %
$-2 \leq v_0^2 < 0$, a level-$k$ $SU(2)$ current algebra $\{J^\pm, J_3\}$ 
appears in the left-mover of the CFT at a complex codimension-1 subspace 
in the Coulomb branch moduli space; $J^+$ has charge $v_0$.
There is an enhanced $SU(2)$ gauge symmetry in the 4d effective theory
(e.g. \cite{AP}, \cite{EW}). The $SU(2)$ Cartan charge is 
\begin{align}
q_v := \frac{(v_0,v)}{(v_0,v_0)} \in \frac{1}{2} \mathbb{Z}
\end{align}
for states whose $U(1)$ charge is $v\in \widetilde{\Lambda}_S^{\vee}$. 

Since only massless matters contribute to the anomaly, we can focus our attention to the BPS states of the 4d ${\cal N}=2$ supersymmetry algebra.
The multiplicities of purely electrically charged BPS states are captured by the new supersymmetric index in the Heterotic worldsheet language:
\begin{align}
    Z_{\mathrm{new}}(\tau,\bar{\tau})
    &= \frac{-i}{\eta(\tau)^{24}} \Tr_{\text{R-sector}}
    \sqbra{
    e^{\pi i F_R} F_R\;
    q^{L_0-c/24} \bar{q}^{\bar{L}_0-\tilde{c}/24}
    }
    \\
    &=: \sum_{\gamma\in G_S}
    \theta_{\widetilde{\Lambda}_S[-1]+\gamma}(\tau,\bar\tau)
    \frac{\Phi_\gamma(\tau)}{\eta(\tau)^{24}},
\end{align}
where the trace is over the internal $(c,\tilde{c}) = (22,9)$ worldsheet CFT
with the right-mover in the Ramond sector, 
and $F_R$ is the zero mode of total $U(1)$ current in the right-mover \cite{BDetal}.
The function $\theta_{\widetilde{\Lambda}_S[-1]}$ is the Siegel theta function for the lattice $\widetilde{\Lambda}_S[-1]$. (See \cite{EW} for more details.)
The integral coefficients $c_\gamma(\nu)$ defined by
\begin{align}
    \frac{\Phi_\gamma(\tau)}{\eta(\tau)^{24}}
    =: \sum_{\gamma \in \gamma^2/2 + \Z} c_\gamma(\nu) q^\nu,
    \qquad c_\gamma(\nu) = 0 \text{  for } \nu < -1,
\end{align}
is relevant to BPS counting.
In particular, when $v\in\widetilde{\Lambda}_S^\vee$ satisfies $[v]\neq 0 \in G_S$ and $-2 < v^2 < 0$, there are $n_\gamma^{V/H}$ BPS vector/half-hyper multiplets of $U(1)$-charge $v$, and the following equality holds:
\begin{align}
    c_{[v]}(v^2/2) = -2n_{[v]}^V + n_{[v]}^H \equiv n_{[v]}^H \bmod 2.   
\end{align}
The leading Fourier coefficients $c_\gamma(\nu)$ for $\gamma \in G_S$, 
where $-1 \leq \nu < 0$, are denoted by $n_\gamma$.\footnote{
    The discriminant group $G_S$ may contain an element $\gamma \neq 0$ with $(\gamma,\gamma)\in 2\Z$. In this article we do not assume that $n_{\gamma}=0$ for such a non-trivial isotropic element $\gamma$ although there is no type IIA dual in the geometric phase when $n_{\gamma}\neq 0$ for such a $\gamma$.
}
The ${\cal N}=2$ 
supersymmetry in $\R^{3,1}$---not more, not less---implies $n_\gamma=-2$ 
for $\gamma = 0 \in G_S$.  
In the following arguments, we implicitly use the integrality of BPS indices $c_\gamma(\nu)$.

\section{$\Lambda_S = U \oplus W$ Cases}

We show that the $SU(2)$ Witten anomaly vanishes if $\Lambda_S$ has 
a decomposition $\Lambda_S = U \oplus W$ and $v_0 \in W^\vee \subset 
\widetilde{\Lambda}_S^\vee$; here, 
the lattice $W$ is even and negative definite and is assumed to have a primitive embedding into $U^{\oplus 2}\oplus E_8[-1]^{\oplus 2}$.

\subsection{Derivation Using also the Discrete Peccei--Quinn Symmetries}
\label{ssec:use-d-int}

In the 4d effective theory, the action of the vector multiplets and the 
mass of the BPS states of the ${\cal N}=2$ supersymmetry 
are determined by the prepotential
\begin{align}
     \mathcal{F} &= \frac{s}{2}(t,t) + \frac{d_{abc}}{3!} t^a t^b t^c 
         - \frac{a_{ab}}{2}t^at^b - b_a t^a 
         - \frac{\zeta(3)}{(2\pi i)^3} \frac{\chi}{2}
    + {\cal O}(e^{2\pi i s}, e^{2\pi i t}),
   \label{eq:prepot}
\end{align}
where $s$ is the 4d axion-dilaton complex scalar, and 
$t \in \Lambda_S \otimes \C$ collectively denotes the Narain moduli; $t$ 
is a set of local coordinates of 
$D(\widetilde{\Lambda}_S) := \P\{ \mho \in \widetilde{\Lambda}_S \otimes \C 
\; | \; \mho^2 = 0, \; (\mho, \overline{\mho}) > 0 \}$, and 
$t = e_a t^a$ is the component description for an integral basis $\{ e_a \}$ 
of $\Lambda_S$. Parameters $d_{abc}$, $a_{ab}$, $b_a$, and $\chi$ reflect 
more information of the internal space than must the lattice $\widetilde{\Lambda}_S$.

Two points in the Coulomb branch moduli space $D(\widetilde{\Lambda}_S)$
should describe the same lattice vertex operator algebra and an identical 
string vacuum, if they are in a common orbit of 
\begin{align}
 \Gamma_S := \left\{ g \in {\rm Isom}(\widetilde{\Lambda}_S) \; 
    | \; g {\rm ~acts~trivially~on~}G_S \right\}. 
\end{align}
As we will review in appendix \ref{ssec:GammaS-n-EMduality}, 
it follows from the unphysical nature of the $\Gamma_S$ action\footnote{
In the language of Heterotic string in the geometric phase, 
with the structure $\Lambda_S = U \oplus W$ and $W \subset E_8[-1]^{\oplus 2}$, 
the $\Gamma_S$ action on the Coulomb branch moduli (Narain moduli) includes 
$+1$ shift of the complex structure of $T^2$, $+(2\pi)^2 \alpha'$ shift of 
$\int_{T^2} B$, and winding gauge transformation of $W$. In the dual Type IIA 
language, $d_{abc}$'s are the intersection numbers of the internal 
Calabi--Yau threefold. We call these $t^a\rightarrow t^a+\delta^a_d$ shifts discrete Peccei--Quinn symmetries in this article.
} that 
\begin{align}
  d_{abc} \in \Z, \qquad a,b,c\in 1,\cdots, \rho.
  \label{eq:d-int}
\end{align}

The vector-valued modular form $\Phi/\eta^{24}$ determines a part 
of $d_{abc}$ \cite{AFGNT-pert, AFGNT-higher, HM, CL, Stieberger}; 1-loop threshold corrections 
to a probe gauge group and a gravitational coupling depend on 
$\Phi/\eta^{24}$, and hence $d_{abc}$'s on some of the coefficients 
$c_\gamma(\nu)$'s. It is known that $d_{abc}$'s would not be integers 
automatically, if $\Phi$ were a generic vector-valued modular form 
of weight $11-\rho/2$ and in the Weil representation of ${\rm Mp}_2\Z$ of $\Lambda_S$ with integer-valued $c_{\gamma}(\nu)$
for small $\nu$'s. So, the property (\ref{eq:d-int}) imposes non-trivial 
conditions on the BPS state multiplicities.

Appendix B.1.3 of \cite{EW} has worked out for a general $W$ 
which part of the parameters $d_{abc}$'s are determined by $\Phi$ and how. 
Integrality of some of those $d_{abc}$'s is translated into \cite{EW}
\begin{align}
   \label{eq:c-integrality}
    \sum_{v \in W^\vee}^{(v,a) > 0}
    c_{[v]}(v^2/2)\; (v,r_1) (v,r_2) (v,r_3) \in 2\Z,
    \qquad {}^\forall r_{1,2,3} \in W,\, {}^\forall a\in W\tensor\R.
\end{align}
We can use this condition to show that the Witten anomaly $\mathrm{WA} \in \Z_2$ vanishes; to do this, let us use $a = -v_0$ for convenience, and set 
$r_1 = r_2 = r_3 = kv_0$. Then the condition reads
\begin{align}
    \label{WAfree:rank1}
    \sum_{v \in W^\vee}^{q_v > 0}
    n_{[v]}^H\; (2q_v)^3 \in 2\Z.
\end{align}
Since $2q_v$ is integral and $(2q_v)^3 \equiv 2q_v \bmod 2$, this condition is nothing but the condition \eqref{eq:WA-vanish} which means the theory is free from Witten anomaly; only neutral states under $U(1)$'s contribute to (\ref{WAfree:rank1}) as stated earlier, which are states corresponding to $v$'s that are parallel to $v_0$. They are indeed massless when the $SU(2)$ is enhanced.

\subsection{An Approach just Using Modurality of $\Phi$}

It is instructive to see how far one can go by using just the property 
that $\Phi$ is an integer-coefficient vector-valued modular form, without 
using the integrality of $d_{abc}$'s. We consider $W$ and 
$v_0$ such that
\begin{align}
    W = \vev{-2n}, \quad n = 1,2,\cdots, \qquad {\rm  and} 
        \quad  v_0 = 2/(2n) e,
\end{align}
where $e$ is the generator of the rank-1 lattice $W$. The charge vector $v_0 = \frac{2}{2n} e$ satisfies 
the condition \eqref{eq:enhancement-condition} with $k = n$ \cite{EW}. In this setup we will see that 
$\mathrm{WA} \equiv 0 \bmod 2$ if $n \Slash{\equiv} 1$ mod 4. 
The integrality of $d_{abc}$'s turned out to be necessary, however, 
at least for some cases with $n \equiv 1$ mod 4.  

The coefficients of $\Phi/\eta^{24}$ have various linear relations among them.
One way to obtain such a relation is to use the fact\footnote{
For references and more explanations, see \S 2.4 of \cite{EW}, for example.
} %
\begin{align}
 \label{eq:phi-Phi}
  \sum_{\gamma \in G_S}\phi_\gamma^{(1/2)} \Phi_\gamma = -2 [\phi^{(1/2)}]_{q^0} E_4 E_6, 
\qquad  
     \sum_{\gamma \in G_S} \phi_\gamma^{(9/2)} \Phi_\gamma =
     -2 [\phi^{(9/2)}]_{q^0} E_4 ^2 E_6
\end{align}
for a holomorphic vector-valued modular form $\phi^{(w)}$ of weight $w=1/2$ 
or $9/2$ associated with the dual representation of ${\rm Mp}_2\Z$. 
Here, $E_4$ and $E_6$ are Eisenstein series 
of weight-4 and 6, respectively. We assume that $\phi^{(w)}$ has a 
non-zero $q^0$ term only in the $\gamma = 0 \in G_S$ component, and 
the coefficient is denoted by $[\phi^{(w)}]_{q^0}$. 
For any given $\phi^{(w)}$, one linear relation on $c_\gamma(\nu)$'s is 
obtained by comparing coefficients of each term in the Fourier series 
expansion. 

Here, we use $\phi^{(1/2)} = \theta_{W[-1]}$ and 
$\phi^{(9/2)} = (\partial^S)^2 \theta_{W[-1]}$, where $\partial^S$ is 
Ramanujan--Serre derivative defined by
\begin{align}
    \partial^S F = \parren{ q\pdiff{}{q} - \frac{w_F}{12} E_2} F, 
\end{align}
where $w_F$ is the weight of a modular form $F$. 
The linear relations (\ref{eq:phi-Phi}) for the 
$q^1$ term are combined to yield \footnote{
    the -3$\delta_{n,1}$ term should be replaced by a little different expression when $G_S$ has a non-trivial isotropic element $\gamma$ with $n_{\gamma}\neq 0$. The arguments in the rest of this section remain valid even in that case.
}
\begin{equation}
  \sum_{v = \ell e/(2n)}
     \left[ \left(\frac{\ell^2}{4n}\right)^2
          - \frac{1}{4}\left(\frac{\ell^2}{4n}\right)
     \right]  c_{[\ell e/(2n)]}(-\ell^2/4n) -3 \delta_{n,1} = -3
  \label{eq:relation}
\end{equation}

Only BPS states with $v \parallel v_0$ contribute to $\mathrm{WA}$, as stated 
earlier. So 
\begin{equation}
    \mathrm{WA} = 
    \frac{1}{2}\sum_{ -n\leq \ell \leq n, \; \ell \; {\rm odd} }^{-2<-\ell^2/2n<0} n_{\ell} \bmod 2,
\end{equation} 
where $n_{\ell} = c_{[v]}(v^2/2)$ with $v = (\ell/2n) e \in W^\vee$. We will use this omitted notation from now on.
When $n$ is even [resp. $n \equiv 3 \bmod 4$], the 
relation \eqref{eq:relation} multiplied by $(8n^2)$ [resp. $(4n^2)$] 
and reduced modulo 2 implies that $\mathrm{WA} = 0 \bmod 2$.
When $n \equiv 1$ mod 4, the relation (\ref{eq:relation}) does not 
yield useful information on whether $\mathrm{WA} \equiv 0$ mod 2 or not. 

For example, when $n=1$, there is no linear relation among $n_\gamma$'s, 
and any $n_1 \in \Z_{\geq -2}$ is allowed when $c_\gamma(\nu)$'s with small 
$\nu$ are required to be integers. In the cases with $n=5$ and $n=9$, 
on the other hand, there are two and three linear relations among $n_\gamma$'s, 
respectively. But they cannot be combined to yield 
$\sum_{0< \ell,  {\rm odd}}^{\ell^2\leq 4n}n_{\ell} \equiv 0$ mod 2.

\section{The Case $\Lambda_S = \vev{+28}$}
\label{sec:deg-28}

Next we consider the case\footnote{
    For any $n \in \N_{>0}$, a primitive embedding $\Lambda_S \hookrightarrow U^{\oplus 3} \oplus E_8[-1]^{\oplus 2}$ exists.
} $\Lambda_S = \vev{+28}$, because this turns out to be minimum degree rank 1 lattice where the theory is possibly anomalous. In this case
\begin{align}
    v_0=2(e_0'+e_4')+14e/28\in\widetilde{\Lambda_S}^{\vee} = (U[-1]\oplus\vev{+28})^{\vee}
   \label{eq:v-for-Wboson}
\end{align}
satisfies
\begin{align}
    (v_0,v_0)/2=-1/2,\ \ 2v_0=0\in G_S
\end{align}
and there can be an $SU(2)$ current algebra. Here $e_0',e_4'$ and $e$ are generators of $U[-1]$ and $\Lambda_S^{\vee}$, respectively. These satisfy $(e_0',e_4')=(e_4',e_0')=-1, (e,e)=1/28$, and other pairs are 0. As for charged matters,
\begin{align}
    v_1=v_0/2 = e_0'+e_4'+7e/28
\end{align}
has Cartan charge 1/2. This is the only relevant charged matter.\footnote{
    Charged matter $v\in \widetilde{\Lambda_S}^{\vee}$ that contributes to Witten anomaly satisfies
    \begin{eqnarray}
        v \parallel v_0, \ \ -2<(v,v)<0,\ \ q_v\in 1/2+\Z.\nonumber
    \end{eqnarray}

    $v_1$ and $-v_1$ are the only elements satisfying these conditions.
} At subspace in the Coulomb branch moduli space where these vectors and matters become massless, the effective field theory is an $SU(2)$ (and $U(1)$ graviphoton) gauge theory with $n_7^H$ half-hyper doublets. Anomaly in this theory is equal to $n_7$ mod 2.

It is not known if there exists a Heterotic compactification with 
$\Lambda_S = \vev{+28}$; no consistency condition is known to rule out 
such a Heterotic compactification, however. So we assume that there is 
one, and end up with a conclusion that modularity of some invariants 
of this compactification and the conditions (\ref{eq:d-int}, \ref{eq:Wall-a}, \ref{eq:Wall-bb}) combined fail to predict that the $SU(2)$ Witten anomaly 
vanishes automatically. This is a clear indication that there are more 
theoretical consistency conditions in the worldsheet CFT than those 
properties we use in this analysis. 

\subsection{Modular Forms $\Phi$ and $\Psi$}

Suppose that there is a Heterotic compactification with $\Lambda_S=\vev{+28}$. 
Then two vector-valued modular forms $\Phi$ and $\Psi$ extract invariants 
of this string vacuum; $\Phi$ has appeared already in this article; 
the other one, $\Psi$, is of weight 25/2, and also in the Weil representation 
of ${\rm Mp}_2\Z$ associated with the lattice $\Lambda_S = \vev{+28}$. 
The modular form $\Psi$ as an invariant of a Heterotic--Type IIA dual vacuum 
appears manifestly already in \cite{HM}; see \cite{EW} for references and treatment of $\Psi$ for a general $\Lambda_S$.
The parameters $d_{abc}$ are determined by the Fourier coefficients of 
both $\Phi$ and $\Psi$, and hence we need to deal with both\footnote{
In the case of $\Lambda_S = U \oplus W$, some of $d_{abc}$'s depend only on 
the Fourier coefficients of $\Phi$, not both. It just happens that 
the condition (\ref{eq:d-int}) for such $d_{abc}$'s is enough to guarantee 
that $\mathrm{WA} \equiv 0$ mod 2. 
} %
to exploit the condition (\ref{eq:d-int}).  

The vector space of vector-valued modular forms of weight $21/2$ and Weil 
representation for $\vev{+28}$ is of 14 dimensions, which we find by 
using the Riemann--Roch theorem \cite{Borch-GKZ}.
There is one linear relation among the 15 parameters $n_\gamma = n_{-\gamma}$ 
with $\pm \gamma \in G_S/(-1)$; $G_S \cong \Z_{28}$,  
and the relation can be used to solve one of them in terms of the others:
\begin{align}
    \label{eq:lin-rel}
    n_9& \; =6n_0+3n_1-13n_2-3n_3+10n_4-n_5-3n_6+n_7+10n_8 \nonumber \\
       & \qquad -n_{10}-3n_{11}+4n_{12}+3n_{13}-3n_{14}.
\end{align}
All the Fourier coefficients $c_\gamma(\nu)$ with $\nu \leq 1$ turn 
out\footnote{
To work out those Fourier coefficients with $\nu \leq 1$ in terms 
of the 14 independent $n_\gamma$'s, and also to derive (\ref{eq:lin-rel}, \ref{eq:chi-deg28}), 
we used holomorphic Jacobi forms of index-14 and weight 4, 6, $\cdots$, 12, 
and 16. See sections 2.4 and 3.1.2 of \cite{EW}.
} %
 to be 
integers as long as all the 14 $n_\gamma$'s on the right-hand side are integers. 
The allowed region of $n_{\gamma}$'s is \cite{EW}
\begin{equation}
    \begin{split}
    \label{region cdn}
    n_{\gamma} \geq 0\ ({\rm for}\  \gamma=0,\cdots, 13),\ \ n_{14}\geq -2,\\
    \chi \leq 2(\rho+1)=4,
    \end{split}
\end{equation}
where $\chi$ is the following combination (the Euler number of the dual 
Type IIA Calabi--Yau)
\begin{align}
    \chi & \; =-300+112n_1-154n_2+182n_4-42n_6+16n_7+210n_8  \nonumber \\
       & \qquad \qquad -14n_{10}+70n_{12}+112n_{13}-40n_{14}.
   \label{eq:chi-deg28}
\end{align}
This allowed range is non-empty. For example, $n_0 = -2$, $n_{12} = 3$, 
$n_7 = n_{10} = n_*$ with $0 \leq n_* \leq 47$, and $n_\gamma = 0$ 
for all other $\gamma$'s. 

Similarly, the vector space of vector-valued modular forms of weight 25/2 
and Weil representation for $\vev{+28}$ is of 17 dimensions, which we 
compute by using the Riemann-Roch theorem. It turns out that the Fourier 
coefficients $c_\gamma^\Psi(\nu)$'s in 
\begin{align}
 \frac{\Psi}{\eta^{24}} = \sum_{\gamma \in G_S} c^\Psi_\gamma(\nu) q^\nu
\end{align}
are parametrized by the 15 $c^\Psi_\gamma(\nu) = c^\Psi_{-\gamma}(\nu)$'s with 
$-1 \leq \nu < 0$, and two more, $c_{\gamma =0}(0)$ and $c_{\gamma = 1}(1/56)$; 
we have seen that\footnote{
Similarly to the case of $\Phi$, now we used holomorphic Jacobi forms 
of index-14 and weight 4, 6,$\cdots$,10, and 14.
} %
 all of $c^\Psi_\gamma(\nu)$ with $\nu \leq 1$ are integers 
as long as all the 17 independent parameters are integers.  
Those 17 extra parameters of the Heterotic vacuum in question cannot be 
arbitrary integers, however. They are subject to the conditions \cite{EW}
\begin{align}
 d_\gamma(\nu) \in 12\Z,  \qquad m_0=m_1=m_2=m_3=m_8=m_{11}=m_{13}=0, 
  \label{eq:Psi-range}
\end{align}
where $d_\gamma(\nu)$'s are the Fourier coefficients of 
$(\Phi E_2 -\Psi)/\eta^{24}$, and the leading coefficients 
($d_\gamma(\nu)$'s with $-1\leq \nu <0$) are denoted by $m_\gamma$. 
$d_0(0)$ and $d_1(1/56)$ are simply denoted by $d_0$ and $d_1$ in 
the following.  

\subsection{The Analysis}

The prepotential ${\cal F}$ (\ref{eq:prepot}) and the gravitational coupling $F_1$ in 4d 
\begin{align}
  F_1 = 24 s + (c_2)_a t^a + {\cal O}(e^{2\pi i s}, e^{2\pi i t})  
 \label{eq:F1}
\end{align}
are determined by $\Phi$ and $\Psi$; a survey of the procedure of computation 
applicable to the $\Lambda_S = \vev{+28}$ case is available in \cite{EW}. 
The result in an appropriate basis is
\begin{equation}
    \label{dandc}
    \begin{split}
   (c_2)_1 =&(- 122 n_0 - 130 n_1+ 46 n_2 - 30 n_3 - 126 n_4 - 8 n_5 + 22 n_6 - 10 n_7\\
    &-128 n_8 + 8 n_{10} + 14 n_{11} - 38 n_{12} -36 n_{13} + 26 n_{14}+6l) 
    \\
    d_{111} 
       =&(- 191 n_0 - 199 n_1+ 97 n_2 - 57 n_3 -237 n_4 - 20 n_5 + 37 n_6 - 19 n_7\\
    &- 236 n_8 +14 n_{10} + 41 n_{11} - 65 n_{12} - 30 n_{13} + 50 n_{14}+21l) 
    \end{split}
\end{equation}
where\footnote{
    By following the arguments in \cite{EW}, one can see that the extra parameters $d_\gamma(\nu)$'s of $\Psi$ determine $(c_2)_a$ and $d_{abc}$ only through at most $\rho$ independent linear combinations. In the case of a lattice $\Lambda_S$ with a given rank $\rho$ and larger $G_S$, there tends to be more independent $d_\gamma(\nu)$'s than $\rho$, as in (\ref{dandc}). We have multiple (mutually non-exclusive) interpretations for what is happening. In the language of dual Type IIA string theory, (a) there may be multiple Calabi-Yau three-folds in a common diffeomorphism class that are distinct in their symplectic/complex structure, (b) there may be multiple different hypermultiplet-moduli tunings of a Calabi-Yau three-fold $X$ so that a singularity develops along the same curve class,  and (c) there are more theoretical constraints on $d_\gamma(\nu)$'s than those discussed in \cite{EW}.}    
\begin{align}
    l=(-3d_0/2-d_1+7m_4+5m_5+3m_6+m_7+2m_9+m_{10}-m_{12})/12.
\end{align}

The property (\ref{eq:d-int}) implies\footnote{
The parameters $(c_2)_1$ and $d_{111}$ have also been worked out in terms of 
the coefficients of $\Phi$ and $\Psi$ in the case of 
$\Lambda_S = \vev{+2}$ \cite{EW} and $\Lambda_S = U$. The property (\ref{eq:d-int}) 
implies $d_0/24 \in \Z$, and hence the absence of the $SU(2)$ Witten 
anomaly in the probe gauge group also in the case of $\Lambda_S=\vev{+2}$ and $U$. 
} %
that $d_0 \in 24\Z$, not just 
$d_0 \in 12\Z$ as in (\ref{eq:Psi-range}). This implies that the probe 
gauge group is free from the $SU(2)$ Witten anomaly 
(see discussions in \cite{EW}), but $n_7$ can be an arbitrary integer, 
so the $SU(2)$ gauge group for the $v_0$ in (\ref{eq:v-for-Wboson}) may 
have non-vanishing Witten anomaly. 

If this Heterotic compactification has a Type IIA dual, and the dual 
is in a geometric phase, then it has a property that (see discussion 
around (\ref{eq:Wall-bb}))
\begin{align}
    (c_2)_1 + 2d_{111} \in 12\Z .      
\end{align}
This only leads to 
\begin{align}
    n_{14}\in 2\Z.
\end{align}
The $SU(2)$ gauge group associated with $v_0$ in (\ref{eq:v-for-Wboson}) 
may have only an even number of half-hypermultiplets in the adjoint 
representation. Consequently there is no restriction on $n_7$ other 
than (\ref{region cdn}). 

The modular nature of the elliptic genus was enough in proving that 
the low-energy gauge theory of a Heterotic compactification is 
free from perturbative anomalies \cite{SW}. By examining the 
$SU(2)$ Witten anomaly in the low-energy effective theory, however, 
we found that there is more consistency conditions in the internal space worldsheet SCFT other than the modularity and integrality\footnote{
    To be rigorous, we have only exploited the integrality of $c_{\gamma}(\nu)$'s and $c_{\gamma}^{\Psi}(\nu)$'s with small $\nu$'s in this article. Note also, the modular invariance of a Heterotic string compactification may be more than the modularity of that of $\Phi$ and $\Psi$.
} 
of the invariants $\Phi$ and $\Psi$ and the unphysical discrete 
Peccei--Quinn shift symmetries $t^a \rightarrow t^a + \delta^a_{\; d}$.

\section*{Acknowledgements}

We thank K. Hori and Y. Tachikawa for useful comments and stimulating 
discussions.
This work is supported in part by the World Premier International Research 
Center Initiative (WPI),   
Grant-in-Aid New Area no. 6003 (YE, YS and TW), 
the FMSP program (YE), the IGPEES program (YS), all from MEXT, Japan.

\appendix

\section{Consequences of the Discrete Peccei-Quinn Shift Symmetries}
\label{ssec:GammaS-n-EMduality}

Here, we think of a Heterotic compactification described in 
section \ref{sec:set-up}. The lattice $\Lambda_S$ is not necessarily 
of the form $U \oplus W$. We neither assume that the $(c,\tilde{c}) = (22,9)$
CFT has a Lagrangian-based description in the Heterotic language, nor 
assume that the Type IIA dual is in a geometric phase. 
By exploiting the fact that the $\Gamma_S$ action on $D(\widetilde{\Lambda}_S)$ 
should be unphysical, we will derive (\ref{eq:d-int}) (cf. \cite{HKTY,Ferrara-duality,deWit,on-quintic}). 

Consider changing the Narain moduli $t^a$ continuously to 
$(t')^a = t^a + \delta^a_d$ for some $d = 1,\cdots, \rho$. 
There is an isometry $g_{(d)} \in {\rm Isom}(\widetilde{\Lambda}_S)$ so that 
$\mho(t') = g_{(d)} \cdot \mho(t)$, so $\vev{v, \mho(t)} = \vev{v', \mho(t')}$
for $v' := (g_{(d)}^{-1})^T \cdot v$ for any $v \in \widetilde{\Lambda}_S^\vee$; 
it is straightforward to verify that this $g_{(d)}$ is in the subgroup 
$\Gamma_S \subset {\rm Isom}(\widetilde{\Lambda}_S)$. Thus, we should have 
arrived at the original vacuum at the end. The BPS mass spectrum should be 
the same before and after this deformation. 

States from F1 Heterotic string have purely electric charges under the 
$(2 + \rho)$ $U(1)$ vector fields in the 4d effective theory; 
the electric charges (simply referred to as charges outside of this appendix) 
$v$ fill the free abelian group $\Lambda_{\rm el} = \widetilde{\Lambda}_S^\vee$. 
States with magnetic charges under the $(2+\rho)$ 
$U(1)$ vector fields arise from NS5-branes and objects of similar kinds; 
the magnetic charges $m$ fill the free abelian group 
$\Lambda_{\rm mg} = \Lambda_S$. 
The Dirac--Zwanziger quantization condition is satisfied, with a canonical 
symplectic form introduced on the free abelian group 
$\Lambda_{\rm el} \oplus \Lambda_{\rm mg} = 
\widetilde{\Lambda}_S^\vee \oplus \Lambda_S$ \cite{AFGNT-pert}.
So, the transformation $g_{(d)}$ on $\widetilde{\Lambda}_S^\vee = \Lambda_{\rm el}$
should be lifted to a consistent action $\tilde{g}_{(d)}$ on 
$\Lambda_{\rm el} \oplus \Lambda_{\rm mg}$, and the spectrum of the central 
charges remain the same before and after the deformations. 

The central charge\footnote{
The mass of BPS states of charge $Q = (v_I, m^I)$ is given by 
$m = |Z|/\sqrt{G_N}$, where $G_N$ is the 4d Newton constant. 
} of the 4d ${\cal N}=2$ supersymmetry algebra 
is given by (e.g., \cite{Ferrara-duality, deWit})
\begin{align}
  Z & \; = e^{K/2} (v_I X^I + m^I F_I), \qquad 
       K = - \ln( i(X^I \overline{F}_I - \overline{X}^I F_I)), \\
  & X^I = X^0(1,\; t^a,\; (t,t)/2+{\cal O}(e^{2\pi is})), \\
  & F_I = X^0(2{\cal F}-(s\partial_s + t\partial_t){\cal F}, \; 
      \partial_{t^a} {\cal F}, \;  -s ), 
\end{align}
where $I, J$ range from 0, $1, \cdots, \rho$, $\rho+1$; 
$a,b = 1,\cdots, \rho$. We use a prepotential of the form 
\begin{align}
 {\cal F} = \frac{s}{2}(t,t) + \frac{d_{abc}}{6} t^a t^b t^c
    - \frac{a_{ab}}{2}t^a t^b - b_a t^a 
   - \frac{\zeta(3)}{(2\pi i)^3} \frac{\chi}{2}
   + {\cal O}(e^{2\pi i s}, e^{2\pi i t}).  \nonumber 
\end{align}
A rationale for assuming this form of ${\cal F}$ even when the Type IIA 
dual is not necessarily associated with a geometric phase is found 
toward the end of this appendix. 

The lift $\tilde{g}_{(d)}$ acting on $(X^A, F_A)^T$ should be 
\begin{align}
  \tilde{g}_{(d)} = \left( \begin{array}{cc} 
       [g_{(d)}]^A_{\; B}  &  0^{AB} \\
       W_{AB} & [(g_{(d)}^{-1})^T]_A^{\; B} 
    \end{array} \right), 
\end{align}
with 
\begin{align}
 W_{AB} = \left( \begin{array}{ccc}
     -\left( 2b_d+\frac{d_{ddd}}{6} \right) &  
     - \left( a_{db} + \frac{d_{ddb}}{2} \right) &  0 \\
    - a_{ad} + \frac{d_{add}}{2} & d_{adb} & 0 \\
      0  & 0 & 0
    \end{array} \right).
\end{align}
For $(\tilde{g}_{(d)}^{-1})^T$ to be interpreted as relabeling of electric 
and magnetic charges $(v_A, m^A) \in \Lambda_{\rm el} \oplus \Lambda_{\rm mg}$, 
\begin{align}
    d_{abd} \in \Z \qquad \qquad{}^\forall a,b,d \in 1,\cdots, \rho, 
\end{align}
and the parameters $a_{ab}$ and $b_a$ should be chosen so that\footnote{
Difference $\Delta a_{ab} \in \Z$ and $\Delta b_a \in \Z$ 
is absorbed by a relabeling of the charges of the form 
$(v'_I, m^{'I}) = (v_I + w_{IJ}m^I, m^I)$ with $w_{IJ} \in \Z$ and $w_{IJ} = w_{JI}$.
} %
\begin{align}
    \label{eq:Wall-b}
    a_{ab} \in \left( \frac{d_{abb}}{2} + \Z \right) \cap \left(\frac{d_{bba}}{2} + \Z\right), \qquad 24b_a + 2 d_{aaa} \in 12\Z. 
\end{align}
For the fractional part of $a_{ab}$ to be determined consistently, 
parameters of the prepotential should also satisfy 
\begin{align}
    d_{abb} + d_{aab} \equiv 0 \quad {\rm mod~} 2, \qquad
    {}^\forall a,b\in 1,\cdots, \rho. 
 \label{eq:Wall-a}
\end{align}
Therefore, both (\ref{eq:d-int}) and (\ref{eq:Wall-a}) must be satisfied
in a Heterotic compactification under consideration. 

In the matching calculation using the Heterotic 1-loop threshold 
corrections to the gauge coupling of a probe level-1 gauge group,\footnote{
In this paragraph, we use notations, jargons, and detailed discussions 
in \cite{EW} without due amount of explanations. Curious readers are 
referred to \cite{EW}.
} %
the parameters $d_{abc}$ and the 1-loop corrected gauge kinetic 
function $f_{({\cal R})}= (s+d'_a t^a)$ of the probe gauge group are determined only 
modulo shift $d_{abc} \rightarrow d_{abc} + [(\delta n_a)C_{bc} + {\rm cyclic}]$
and $f_{({\cal R})} \rightarrow f_{({\cal R})} + (\delta n_a)t^a$ 
(cf \cite{KL}).\footnote{
$C_{ab}$ is the intersection matrix of $\Lambda_S$ presented in an
integral basis $\{e_a\}$ of $\Lambda_S$. 
} %
Because the parameters $d'_a$ are regarded as part of the cubic term 
coefficients of the prepotential in the Coulomb branch of the probe 
gauge group, it follows that $d'_a \in \Z$. This indicates---based on 
the presentation in \S 3.1.3 of \cite{EW}---that the property (\ref{eq:d-int}) 
is equivalent to the integrality of $d^{(P)}_{abc}$ in the cubic polynomial 
$P_3(t) =: d^{(P)}_{abc}t^a t^b t^c$ obtained in a Heterotic 1-loop computation.
This subtle chain of logic is implicit in sections \ref{ssec:use-d-int} 
and section \ref{sec:deg-28}.

A rationale for the form of prepotential (\ref{eq:prepot}) is the following. 
Think of the dual Type IIA compactification, and then the third derivative 
of the prepotential ${\cal F}$ should be equal to the three-point function 
of the corresponding A-model, regardless of whether the Type IIA 
compactification is associated with a geometric phase. 
Due to the unphysical nature of Peccei-Quinn shift 
$t^i_X \rightarrow t^i_X + \delta^i_{\; j}$, where 
$(t_X^{i=1,\cdots,\rho+1}) = (t^{i=a=1,\cdots, \rho}, s)$ are the special (flat) coordinates, 
the three point functions should be of the form of 
\begin{align}
  \kappa_{ijk} + \sum_{\beta_{\neq 0} \in L} K_{\beta,ijk} e^{2\pi i \vev{\beta,t_X}}
\end{align}
for some rank-$(\rho+1)$ free abelian group $L$; $\kappa_{ijk}$ and $K_{\beta,ijk}$
are constant parameters. Thus, ${\cal F}$ is a polynomial at most cubic in 
the special coordinates $t_X^i$, besides the exponential terms as 
in (\ref{eq:prepot}). Presumably it is possible to translate this argument 
into the language of Heterotic string.

The prepotential (\ref{eq:prepot}) is not in the most general form one expects from the argument above.
The structure of the $s$-linear term in the cubic part $(\kappa_{ijk}/6) t_X^it_X^jt_X^k$ in ${\cal F}$ is due to that of the tree-level gauge kinetic function of the $(\rho+2)$ 4d vector fields.
The absence of $s^2$ terms and $s^3$ terms in the cubic part, as well as the $s$-independence of the quadratic part $-(a_{ij}/2)t_X^i t_X^j$ and the linear part $-b_i t_X^i$ in (\ref{eq:prepot}) is explained by the fact that the combination $v_I X^I$ in an appropriate frame should be proportional to the right-moving momenta in the Heterotic string.
When a Type IIA dual exists, and is in a geometric phase, all of those structures are translated into the following properties: (i) the fiber K3 class divisor $D_s$ of a Calabi--Yau three-fold with a regular K3-fibration satisfies $D_s^2 =0$, and (ii) $D_s \cdot D_a \cdot D_b = C_{ab}$ \cite{K3-fiber}; (iii) we know in a geometric phase \cite{HKTY, on-quintic, DR} that $b_{(\rho+1)} + \Z = (24)^{-1} \int_X c_2(TX) D_s + \Z = \Z$ and $a_{a(\rho+1)} + \Z = 2^{-1} D_a \cdot D_s \cdot D_s +\Z = \Z$ \cite{K3-fiber}.     

\vspace{5mm}

If a Type IIA dual exists, and is in a geometric phase, 
the property (\ref{eq:d-int}) is a trivial statement that 
all the triple divisor intersection numbers are integers. The property 
(\ref{eq:Wall-a}) is also known as one of the necessary conditions 
for a real 6-manifold \cite{Wall}. In a geometric phase, it is known 
\cite{HKTY, BCOV-1} that the parameters $b_a + \Z$ are equal to 
$(c_2)_a/24 + \Z$ appearing in the gravitational coupling (\ref{eq:F1}). 
So the second property in (\ref{eq:Wall-b}) is read as 
\begin{align}
(c_2)_a + 2d_{aaa} \in 12\Z, 
   \label{eq:Wall-bb}
\end{align}
which is also one of the necessary conditions for a real 6-manifold \cite{Wall}. 
The authors have not been able to build an argument 
relating the fractional part of the parameter $b_a$ in ${\cal F}$ 
and $(c_2)_a/24$ for the coefficient $(c_2)_a$ in (\ref{eq:F1}) 
without assuming that a Type IIA compactification is in a geometric 
phase.\footnote{
We mean to use only the following (i--iii) as assumptions: (i) the internal 
``space'' is described by $N=(2,2)$ compact unitary SCFT with 
$(c,\tilde{c})=(9,9)$, (ii) all the NS--NS sector states have 
integral charges under the left-mover $U(1)$ and the right-mover $U(1)$, 
and (iii) there exists one spectral flow in the left mover, and also one 
in the right mover.  
}\raisebox{5pt}{,}\footnote{
    The set of conditions (\ref{eq:d-int}, \ref{eq:Wall-a}, \ref{eq:Wall-bb}) is not just a necessary condition, but also a sufficient condition for a real 6-manifold to exist \cite{Wall}. So, filling this logical gap is also vital in claiming that all the Type IIA compactifications in the previous footnote are associated with some Calabi-Yau threefold. 
    }  
Despite this caveat, it is tempting to include (\ref{eq:Wall-bb}) 
as one of the properties of a Heterotic compactification 
introduced in section \ref{sec:set-up}.


\begin{thebibliography}{9}

\bibitem{SW}
%
%
%
A.~Schellekens and N.~Warner,
``Anomalies, Characters and Strings,''
Nucl. Phys. B \textbf{287} (1987), 317.
%
%
W.~Lerche, A.~Schellekens and N.~Warner,
``Lattices and Strings,''
Phys. Rept. \textbf{177} (1989), 1.
%
\bibitem{inflow}
%
M.~B.~Green, J.~A.~Harvey and G.~W.~Moore,
``I-brane inflow and anomalous couplings on d-branes,''
Class. Quant. Grav. \textbf{14} (1997), 47-52
[arXiv:hep-th/9605033 [hep-th]].
%

%
\bibitem{brane-CS}
%
M.~R.~Douglas,
``Branes within branes,''
NATO Sci. Ser. C \textbf{520} (1999), 267-275
[arXiv:hep-th/9512077 [hep-th]].
%
Y.~E.~Cheung and Z.~Yin,
``Anomalies, branes, and currents,''
Nucl. Phys. B \textbf{517} (1998), 69-91
[arXiv:hep-th/9710206 [hep-th]].
%
R.~Minasian and G.~W.~Moore,
``K theory and Ramond-Ramond charge,''
JHEP \textbf{11} (1997), 002
[arXiv:hep-th/9710230 [hep-th]].
%
\bibitem{W} E. Witten,
``An SU(2) anomaly,''Phys.Lett. 117B(1982)324-328.
%
%
%
\bibitem{BDetal}
%
T.~Banks, L.~J.~Dixon, D.~Friedan and E.~J.~Martinec,
``Phenomenology and Conformal Field Theory Or Can String Theory Predict the Weak Mixing Angle?,''
Nucl. Phys. B \textbf{299} (1988), 613-626.
%
%
T.~Banks and L.~J.~Dixon,
``Constraints on String Vacua with Space-Time Supersymmetry,''
Nucl. Phys. B \textbf{307} (1988), 93-108.
%
J.~Lauer, D.~Lust and S.~Theisen,
``Supersymmetric String Theories, Superconformal Algebras and Exceptional Groups,''
Nucl. Phys. B \textbf{309} (1988), 771-790.
%
\bibitem{AFGNT-pert}
I.~Antoniadis, S.~Ferrara, E.~Gava, K.~Narain and T.~Taylor,
``Perturbative prepotential and monodromies in N=2 heterotic superstring,''
Nucl. Phys. B \textbf{447} (1995), 35-61
[arXiv:hep-th/9504034 [hep-th]].
%
\bibitem{HM}
%
J.~A.~Harvey and G.~W.~Moore,
``Algebras, BPS states, and strings,''
Nucl. Phys. B \textbf{463} (1996), 315-368
[arXiv:hep-th/9510182 [hep-th]].
%
\bibitem{AP} I. Antoniadis, H.Partouche,
``Exact monodromy group of ${\cal N}=2$ heterotic soperstring,''
Nucl. Phys. B \textbf{444} (1996), 470-488.
%
\bibitem{EW} Y. Enoki, T. Watari,
``Modular forms as classification invariants of 4D N=2 Heterotic-IIA dual vacua''
arXiv:1911.09934v2
%
%
\bibitem{AFGNT-higher}
%
I.~Antoniadis, E.~Gava, K.~Narain and T.~Taylor,
``N=2 type II heterotic duality and higher derivative F terms,''
Nucl. Phys. B \textbf{455} (1995) no.1-2, 109-130
[arXiv:hep-th/9507115 [hep-th]].
%
\bibitem{CL}
%
G.~Lopes Cardoso, G.~Curio and D.~Lust,
``Perturbative couplings and modular forms in N=2 string models with a Wilson line,''
Nucl. Phys. B \textbf{491} (1997), 147-183
[arXiv:hep-th/9608154 [hep-th]].
%
\bibitem{Stieberger}
%
S.~Stieberger,
``(0,2) heterotic gauge couplings and their M theory origin,''
Nucl. Phys. B \textbf{541} (1999), 109-144
[arXiv:hep-th/9807124 [hep-th]].
%
\bibitem{Borch-GKZ}
%
R. Borcherds, ``The Gross--Kohnen--Zagier theorem in higher dimensions,'' 
Duke Math. J. {\bf 97} (1999) 219.
%
\bibitem{HKTY}
%
S.~Hosono, A.~Klemm, S.~Theisen and S.~Yau,
``Mirror symmetry, mirror map and applications to complete intersection Calabi-Yau spaces,''
AMS/IP Stud. Adv. Math. \textbf{1} (1996), 545-606
[arXiv:hep-th/9406055 [hep-th]].
%
\bibitem{Ferrara-duality}
%
A.~Ceresole, R.~D'Auria, S.~Ferrara and A.~Van Proeyen,
``Duality transformations in supersymmetric Yang-Mills theories coupled to supergravity,''
Nucl. Phys. B \textbf{444} (1995), 92-124
[arXiv:hep-th/9502072 [hep-th]].
%
\bibitem{deWit}
%
B.~de Wit, V.~Kaplunovsky, J.~Louis and D.~Lust,
``Perturbative couplings of vector multiplets in N=2 heterotic string vacua,''
Nucl. Phys. B \textbf{451} (1995), 53-95
[arXiv:hep-th/9504006 [hep-th]].
%
%
\bibitem{on-quintic}
%
I.~Brunner, M.~R.~Douglas, A.~E.~Lawrence and C.~Romelsberger,
``D-branes on the quintic,''
JHEP \textbf{08} (2000), 015
[arXiv:hep-th/9906200 [hep-th]].
%
\bibitem{KL}
%
V.~Kaplunovsky and J.~Louis,
``On Gauge couplings in string theory,''
Nucl. Phys. B \textbf{444} (1995), 191-244
[arXiv:hep-th/9502077 [hep-th]].
%
%
\bibitem{K3-fiber}
%
P.~S.~Aspinwall and J.~Louis,
``On the ubiquity of K3 fibrations in string duality,''
Phys. Lett. B \textbf{369} (1996), 233-242
[arXiv:hep-th/9510234 [hep-th]].
%
\bibitem{DR}
%
D.~Diaconescu and C.~Romelsberger,
``D-branes and bundles on elliptic fibrations,''
Nucl. Phys. B \textbf{574} (2000), 245-262
[arXiv:hep-th/9910172 [hep-th]].
%
\bibitem{Wall}
%
C. T. Wall, ``Classification problems in differential topology V. 
On certain 6-manifolds,'' Inv. Math. {\bf 1} (1966) 355. 
%
%
\bibitem{BCOV-1}
%
M.~Bershadsky, S.~Cecotti, H.~Ooguri and C.~Vafa,
``Holomorphic anomalies in topological field theories,''
AMS/IP Stud. Adv. Math. \textbf{1} (1996), 655-682
[arXiv:hep-th/9302103 [hep-th]].


\end{thebibliography}
\end{document}